# Marine Microplastics and Infant Health*

Xinming Du

Shan Zhang

Eric Zou

October 2024

## Abstract

A century of plastic usage has led to an accumulation of plastic waste in waterways and oceans. Over time, these wastes break down into particles smaller than 5 microns – or "microplastics" – which can infiltrate human biological systems. Despite decades of research into this emerging source of environmental pollution, there is a paucity of direct evidence on the health impacts of microplastics exposure at a population scale. This paper reports the first empirical link between in-utero microplastic exposure and adverse birth outcomes. Our analysis is based on a compiled dataset of 3 million births that occurred in coastal areas of 15 countries spanning four continents, which we merge with a novel remote-sensing measurements of marine microplastic concentrations. We show that in-utero exposure to microplastics, particularly during the third trimester of pregnancy, leads to a significant increase in the likelihood of low birth weight. A doubling of exposure increases low birth weight hazard by 0.37 per 1,000 births, which implies over 205,000 cases per year globally can be attributed to microplastic exposure. We further show that aerosolization – whereby microplastic particles become airborne and inhalable due to seawater evaporation – is an important pathway for health impact, a challenge that is likely to escalate as ocean temperatures continue to rise.

* Du: Department of Economics, National University of Singapore (email: xdu@nus.edu.sg); Zhang: Department of Economics, Old Dominion University (email: s2zhang@odu.edu); Zou: Ross School of Business, University of Michigan and NBER (email: ericzou@umich.edu). We thank Maureen Cropper, Tatyana Deryugina, Hannah Druckenmiller, Gabriel Englander, Julian Reif, Chris Ruf, Wolfram Schlenker, Jeff Shrader, Andreas Thurnherr, and participants at various conferences and seminars for helpful comments. Raphaël Pérot provided excellent research assistance. All errors are our own.

# 1. Introduction

Plastic has gained exponential popularity since its invention in the 1900s. It was originally introduced for specialized uses such as in the production of electrical insulator and automotive parts, but the versatility of the material – its durability, light weight, and low cost – soon proved advantageous in a variety of applications, ranging from the lightest everyday uses such as garment making, to the heaviest industrial applications such as aviation (Andrady and Neal, 2009). Plastics production reached 400 million tons in 2022, with packaging being the largest market that accounts for approximately 40 percent of this total amount (Geyer, Jambeck, and Law, 2017).

The environmental impact of plastic wastes has long been debated. Most plastics do not decompose in the same way organic materials do. Instead, they *fragment*, breaking down into smaller and smaller pieces over time. This results in microplastics, tiny plastic particles that can persist in the environment indefinitely (Rillig, 2012). The technology of waste plastic composts and recycling are still immature, leading to an estimated 80 percent of waste ending up in landfills or the natural environment. Much of these wastes are mismanaged: they are either littered or inadequately disposed, with an estimated 4.8 to 12.7 million metric tons ending up in rivers and oceans (Jambeck et al., 2015). Tiny particles may eventually become airborne – a prediction that we will test in this paper – under the right environmental conditions and thus end up in the air that we breathe (Prata, 2018).

Decades of research have accumulated a substantial body of knowledge on emission sources of marine microplastics, their ambient presence and fate, and exposure risks of human and other organisms. In contrast, the availability of data and direct evidence on the health effects remain virtually absent to this date (Lim, 2021; Landrigan et al., 2023). In this paper, we provide the first global-scale estimate of the causal effect of microplastic exposure on infant health. This research addresses three main challenges. First, measurements of marine microplastic pollution are sparse, with most available datasets being cross-sectional. To overcome this, we use a space radar-based remote sensing method that detects microplastic presence on the ocean surface by observing the wind-dampening effect on surface roughness (Evans and Ruf, 2021). This new technique correlates well with in-situ sampling data and provides a longitudinal measurement for microplastics at a high temporal frequency, which is crucial for health analyses that require precise measurement of individual exposure differences.

Second, microplastic concentration in a given location might be correlated with various environmental and economic factors – such as air temperature or water quality – that independently influence health outcomes. We develop an oceanography transport model that leverages idiosyncratic day-to-day variability in ocean current motions to predict how microplastics from the open sea (such as areas over 200 km from the coast) can be expected to reach coastal locations over time. This method helps tease out portions of coastal microplastic exposure that are quasi-random – those that are likely independent to any of the coastal-local confounding factors, which allows us to cleanly isolate out the causal impact of microplastics.

Third, detecting the health effects of microplastic exposure can be challenging due to the complexity of individuals' environmental exposure and the potential delay in health responses. We focus on infant health as mothers are generally stationary during pregnancy, allowing for more accurate exposure measurement (Currie, 2011). Unlike adults, infants do not have a long, complex history of environmental exposure, making econometric modeling of health outcomes and environmental risks—such as in-utero microplastic exposure—relatively straightforward. We compile individual-level data on birth weight from multiple data sources including the Demographic and Health Survey for developing countries, and administrative birth records from Brazil, Mexico, and the United States. Our compiled data includes 3 million individual births from 15 different countries, covering coastlines across four continents. The large sample size and wide geographic scope ensure that our estimates are statistically robust and representative of multiple regions worldwide. We use incidents of low birth weight (infants weighing less than 2,500 grams at birth) as our main infant health marker, a primary measure of infant health that has been linked to environmental stressors (Currie and Almond, 2011).

We find robust evidence that in-utero exposure to microplastics significantly increases the likelihood of low birth weight (LBW) among mothers who lived within 200 km of the coastal line. In our pooled sample, each doubling of microplastic concentration during pregnancy increases LBW incidents by 0.37 per 1,000 births. This effect is primarily driven by exposure during the second and third trimesters. In "placebo" exercises, we show that the same econometric model indicates no impact of microplastic exposure during the preconception and postpartum periods. The effect size we find is significant, suggesting that microplastic variability accounts for about 2 percent of all low birth weight births. This rate ranges from approximately 1.2 percent in the USA to over 8 percent in certain developing countries in Africa and South Asia. Scaled linearly, our estimate implies that about 205,800 of LBW coastal births annually worldwide can be attributed to microplastic exposure.

We explore possible underlying mechanisms, considering two general pathways: water and air. One intuitive channel of impact is the accumulation of microplastics in water and their subsequent consumption through microplastic-contaminated seafood. We leverage two different data sources from retail scanner data from the US and global Automatic Identification System fishing vessels tracking data to build proxies for seafood access. Using both of these proxies, we found no statistically significant evidence that areas with higher seafood access are associated with a larger health effect.

We next turn to the atmospheric pathway. As discussed above, plastic waste can break down into fragments small enough to become aerosolized. Our data shows a consistent pattern: variations in microplastic levels strongly predict the concentration of coastal aerosol particulate pollution, as measured by an independent satellite instrument. This relationship intensifies with higher rates of evaporation in coastal regions. Combining this with established evidence that fine particulates can penetrate multiple organs via respiratory invasion, and specific evidence on the impact of in-utero exposure to ambient particulate matter on low birth weight outcomes, our data suggests aerosolization is a potentially important mechanism through which microplastic pollution affects infant health.

To our knowledge, this paper is the first to establish a direct link between ambient microplastic exposure and health using large-scale data. Our findings corroborate decades of research across multiple fields warning about the potential risks of microplastic exposure. We are guided by knowledge from atmospheric and oceanic sciences that detect microplastics through remote-sensing techniques, from environmental toxicology that predicts the impact of microplastics on organisms, and from medicine that tests the health impacts of microplastics (See our review of current body of knowledge in Section 2). Our results support important hypotheses from prior literature, such as the relevance of microplastic aerosolization. We offer a much larger sample size for studying birth impacts, extending beyond the existing medical evidence that often relies on small sample clinical observations (Amereh et al., 2022).

Our findings on microplastic aerosolization highlights a new, pervasive source of airborne particulate pollution and raise important questions about their broader impacts. There is extensive scientific evidence on the health effects of particulate matter (PM) pollution, which has been linked to adverse outcomes across cardiovascular, respiratory, and mental health domains (e.g., Landrigan et al., 2018). This evidence has motivated regulatory actions on PM globally. Existing research has primarily attributed PM to direct emissions or secondary formations from burning and combustion processes. Our study, however, identifies microplastics as an additional source of ambient PM emanating from the waste sector. The extensive range of impacts and the social costs associated with microplastic-derived PM require further investigation.

From a policy perspective, our research emphasizes the natures in which microplastic pollution is a global challenge. Microplastic particles travel long distances: in our ocean current modeling analysis, a coastal area's microplastic variability can be explained by conditions hundreds of kilometers away in the open sea, which are themselves influenced by emission sources far away. On the health side, the adverse impacts of microplastics appear to be far more widespread than previously appreciated. We find significant effects in both the USA data and in developed countries. The fact that the low birth weight effects do not appear through a seafood intake channel but through a broader atmospheric channel means population exposure is less selective than previously thought. Micron-sized particles are difficult to filter out and can easily penetrate indoor environments, rendering normal protective measures less effective (Chen and Zhao, 2011). Our findings highlight the interconnected nature of the global plastic problem—emissions from one site affect distant areas—and underscore the importance of global plastic waste management and reduction programs. The downstream benefits of such initiatives are currently underappreciated but have the potential to be enormous (Borrelle et al., 2020; He et al., 2023).

Our paper also contributes to the broad literature on in-utero shocks on health outcomes (e.g. Almond, 2006; Almond and Mazumder, 2011; Mansour and Rees, 2012; Currie and Rossin-Slater, 2013; Stearns, 2015; Rangel and Vogl, 2019; Currie et al., 2022). We provide one of the first studies showing ocean pollution affects early-life health. Related to our study, Guimbeau et al. (2024) found ocean salinity affects childhood heights and weights through agricultural intensity and land use changes. Armand and Taveras (2021) documented ocean acidification increases neonatal mortality and affects parental investment on child development, and the channel lies in maternal malnutrition. These two papers focus on ocean chemical composition that is primarily changed by long-term climate change. In contrast, our focus on ocean microplastics has daily variation due to changes in anthropogenic sources and high-frequency ocean currents. To address the endogeneity concern, we explore ocean current direction and speed and construct transported microplastics exposure. This proposed identification strategy is novel and provides a framework for future ocean studies.

Section 2 surveys current body of knowledge on microplastics. Section 3 describes data sources and methods. Section 4 presents the results. Section 5 concludes.

# 2. Current Knowledge

## 2.1 Emission Sources

Since the 1950s, when synthetic organic polymers (plastics) became widely used, global plastic production has surged to 322 million metric tons annually by 2015. About half of this plastic is used for disposables and packaging, with 40% of plastic waste not being properly managed through recycling or landfill facilities. Annually, an estimated 4.8 to 12.7 million metric tons of this waste enters the oceans as both plastic litter and microplastic (MP) particles (Worm et al., 2017).

Microplastics are classified into two types: primary, which are intentionally small-sized plastics used in products like cosmetics, cleaning agents, and medical supplies; and secondary, which result from the breakdown of larger plastic items due to sunlight, physical wear, and biological degradation (Cole et al., 2011). The primary sources of microplastic pollution in the oceans stem from both land-based and ocean-based activities. Land-based sources account for approximately 80% of the microplastics found in the marine environment (Li et al., 2016). These plastics mainly originate from populated or industrial areas characterized by significant littering, usage of plastic bags, and improper waste disposal. For example, coastal recreational activities frequently lead to both floating and beached plastic debris, especially in regions like the northern South China Sea (Lee et al., 2013). Additionally, manufacturing mishaps can also lead to plastic debris being spilled and carried to beaches (Redford et al., 1997). Moreover, plastics reach the oceans via rivers and wastewater systems that transport them from cities and towns (Browne et al., 2010; Cole et al., 2011). Heavy rainfall and extreme weather events, such as hurricanes, can further increase the movement of these plastics from land to sea (Barnes et al., 2009). Ocean-based activities contribute the remaining 20% of marine plastic debris (Li et al., 2016). The most significant contributor in this category is commercial fishing, which loses about 640,000 tons of fishing gear each year. This includes items like nets and lines that continue to trap and harm marine life, a phenomenon known as "ghost fishing" (Good et al., 2010). Figure 1(a) provides a summary of various microplastic sources.

Several studies have aimed to quantify the total number and weight of microplastic particles in marine environments using various methodologies. For example, Jambeck et al. (2015) estimated that populations living within 50 km of coastlines worldwide generate over 8 million tonnes of mismanaged plastic waste annually, much of which enters the oceans. Van Sebille et al. (2015) introduced a statistical framework to standardize a global dataset of plastic marine debris collected via surface-trawling plankton nets. By integrating this data with three different ocean circulation models, they spatially interpolated observations to estimate that between 15 and 51 trillion plastic particles, weighing as much as 236,000 tons, have accumulated in marine systems by 2014. Similarly, Eriksen et al. (2014) used methods including surface net tows and visual surveys in various global marine regions, and reported finding a minimum of 5.25 trillion plastic particles, totaling 268,940 tons, in the world's oceans. These findings emphasize the significant extent of marine microplastics pollution. It is worth noting that most of these studies that describe global microplastic patterns uses cross-sectional measurements. As we will detail further below, for the purpose of our research, it is crucial to consider not only the spatial variability of microplastic exposure but also the temporal variation – where within the same area, births that occur at different times have different in-utero microplastic exposure. This motivates us to use the new spaceborne global measurements of microplastic distribution developed by Evans and Ruf (2021).

## 2.2 Ambient Presence

Microplastics have become a prevalent environmental concern, with their presence detected in various ambient environments, including water bodies, soil, and the atmosphere.

These tiny particles, generally defined as being less than 1 mm in size, enter the marine environment through various pathways, including coastal tourism, fishing activities, marine vessels, industries, and the breakdown of larger plastic debris (Cole et al., 2011). Notably, high concentrations of microplastics have been reported floating in remote ocean areas, particularly in convergence zones of subtropical gyres (Cozar et al., 2014). Furthermore, microplastics have been discovered in deep-sea sediments, marking their presence in even the most remote marine environments (Cauwenberghe et al., 2013).

The distribution of microplastics is not confined to marine settings. Research shows that these particles are also pervasive in freshwater systems and soil. For example, Lake Hovsgol in Mongolia, a large mountain lake, has shown contamination (Free et al., 2014). Rodrigues et al. (2018) discovered microplastics distributed in both water and sediments of the freshwater system in Antua River, Portugal. Studies have also reported the presence of plastic in soil (Rillig 2012; Lwanga et al., 2016; Li et al., 2019). Rillig et al. (2017) discovered that microplastic particles can be transported from the soil surface down through the soil profile by earthworms. This movement has implications for the exposure of other soil organisms, the duration of microplastics residence at greater depths, and the potential eventual arrival of microplastics in groundwater.

Microplastics have been found even in atmosphere, indicating that their small size allows them to be airborne and inhaled, potentially causing respiratory issues (Prata 2018; Liu et al., 2019). This airborne nature of microplastics highlights their ability to infiltrate even remote areas, affecting both urban and rural settings. Recent studies have proposed the mechanisms by which microplastics can transfer from aquatic environments to the atmosphere. Lehmann et al. (2021) demonstrates that microplastics can be ejected into the atmosphere by raindrops; Herb et al. (2023) showed that microplastics can be emitted through bubble bursting in water bodies. This aerosolization process is influenced by the concentration in water and their particle size. Additionally, atmospheric studies have shown that microplastics can be transported over significant distances by air currents, with sea spray and atmospheric movements playing a crucial role in this global distribution (Allen et al., 2020; Caracci et al., 2023). See Figure 1(c) for an illustration.

These findings collectively underscore the pervasive nature of microplastics across various compartments of the ambient environment, emphasizing their potential to be ingested by marine life and subsequently enter human food chains and life.

## 2.3 Environmental Exposure

Microplastics pose a significant environmental threat to organisms in both terrestrial and marine ecosystems. In terrestrial ecosystems, microplastics are found to impact the biophysical properties of soil, suggesting that its widespread contamination may have negative consequences for plant performance and, consequently, for agroecosystems and terrestrial biodiversity (Machado et al., 2019).

In marine ecosystems, microplastics can accumulate and move around in the bodies of invertebrates such as corals and shellfish (Cole et al., 2011) and negatively impact the health and feeding behavior of zooplankton, which serve as intermediary species that transfer energy in the ecological food chain (Cole et al., 2013). Microplastics also serve as vectors for chemical pollutants such a pyrene, which get absorbed by mussels and concentrated in their tissues (Avio et al., 2015). The ingestion of plastics has also been observed in marine animals such as seabirds, turtles, crustaceans, and fish, leading to adverse effects such as gastrointestinal blockages and disruption of normal feeding and reproductive behaviors (Derraik, 2002; Ryan et al., 2009; Cole et al., 2011; Lozano and Mouat, 2009; van Franeker et al., 2011; Trevail et al., 2015; Schuyler et al., 2014). See Figure 2(b) for an illustration. However, the comprehensive impact of microplastics, including their role in contaminant transfer within the marine food web and potential health implications for human consumption of affected seafood, requires further exploration (Carbery et al., 2018).

## 2.4 Health Effects

Existing evidence showing the direct health consequences of microplastics exposure is very limited (Lim, 2021). To the best of our knowledge, only three papers demonstrate a correlation with health outcomes. All of them use small sample sizes with limited spatial coverage, short study periods, and sampling schemes that may not fully represent the general population.

Specifically, Yan et al. (2022) investigated the correlation between microplastics exposure and inflammatory bowel diseases (IBD). The authors recruited two groups of participants in Nanjing, China: 50 healthy participants and 52 IBD patients, with other non-IBD characteristics similar between groups. Fecal samples revealed microplastics in both groups, with higher concentrations, wider distribution of sizes, and more polyethylene terephthalate microplastics in the IBD group. Microplastics concentrations are positively correlated with HBI score and Mayo score, both capturing IBD activity. The authors also used questionnaires to collect basic information and found that participants who drank bottled water, consumed takeaway food, or had higher dust exposure at work exhibited higher microplastics concentrations in fecal samples, suggesting sources of microplastics exposure through digestion.

The second study by Baeza and Martinez (2022) detected airborne microplastics in human respiratory systems. The authors recruited 44 patients in Spain and found microplastics in the lower airways of 30 patients. Higher microplastics levels were observed in older age groups, active smokers, and those with high-risk occupations. Combining microplastics samples with X-ray diagnoses, higher microplastics levels are correlated with radiological abnormalities, increased pathological microbial growth, and reduced forced vital capacity, indicating potential respiratory health risks associated with microplastics exposure.

The third paper by Amereh et al. (2022) studied 43 pregnant women in Iran, focusing on how microplastics are associated with birth outcomes. Placenta samples collected within 10 minutes after delivery showed microplastics in all 13 pregnancies with intrauterine growth restriction, compared to only 3 out of 30 normal pregnancies, suggesting a positive correlation between abnormal pregnancies and higher microplastics levels. Additionally, there was a negative correlation between microplastics concentration and birth weight, length, head circumference, and 1-minute APGAR score.

In contrast, a larger number of papers have documented the risks of microplastics exposure by identifying microplastics in food, water, and the environment without directly measuring health outcomes. Microplastics have been found in various marine species and seafood across different regions, including mainland China (Fang et al., 2019; Li et al., 2022; Pan et al., 2022), Hong Kong (Lam et al., 2023), Taiwan (Lu et al., 2021), Thailand (Akkajit et al., 2022), India (Chen et al., 2021; Selvam et al., 2021), Indonesia (Luqman et al., 2021), Iran (Akhbarizadeh et al., 2019), Nigeria (Mahu et al., 2023), Mexico (Martinez-Tavera et al., 2021), Italy (Squillante et al., 2023), Montenegro (Boskovic et al., 2023), Portugal, and Atlantic Ocean (Barboza et al., 2020). Affected species include and are not limited to clams, shellfish, latus, and punctatus, and cover both wild and commercially caught species. These findings suggest potential human health risks due to contaminated seafood consumption. Among them, Luqman et al. (2021) also detected microplastics in coastal residents' stool samples, although they did not analyze other health consequences. Motivated by the seafood contamination, our study tests whether seafood consumption is a potential mechanism through which marine microplastics could affect human health.

Moreover, microplastics have also been detected in water bodies in the US, China, Austria, Saudi Arabia (Liao et al., 2020), and India (Khaleel et al., 2023). They potentially expose humans through direct contact or contaminated drinking water sources. Microplastics also exist in the air as components of particle pollution affecting human health. Studies have assessed airborne microplastics in the UK (Wright et al., 2020) and Iran (Dehghani et al., 2017). Additionally, microplastics have been found in human lung tissue by Amato-Lourenco at al. (2021) and sputum samples by Huang et al. (2022), although direct effects on human health or respiratory malfunctions were not confirmed. Our paper fills in the gap by investigating the direct links between inland water and air pollution and marine microplastics exposure, and how these factors affect birth outcomes.

# 3. Data and Methods

## 3.1 Data Sources

**Marine Microplastics.** Our main microplastics measurement is based on version 1.0 of the CYGNSS Level 3 data product (Evans and Ruf, 2021). This method uses spaceborne bistatic radar measurements and exploits the reduction of ocean surface's roughening in response to wind when the ocean surface is present with surfactants that act as tracers for microplastics. In other words, a dampening between the empirical relationship between sea surface roughening and wind indicates the presence of a higher concentration of microplastics in the area. This data comes with a resolution of daily by 0.25 degrees latitude/longitude grid. For each coastal birth in our dataset, we assign microplastic exposure based on the concentration observed in the nearest coastal grid to the birth location during the pregnancy.

Figure 2, panel (a), plots the spatial distribution of microplastic concentration. Due to the angular limit of space radar instruments, measurements are available between -37 to 37 degrees latitude. In panel (b), we report aggregated weekly trends in microplastic concentration in seven

areas where our birth data are concentrated. One notable feature from the chart is that microplastic concentrations exhibit different and quite idiosyncratic trends in various parts of the world. This is useful in our econometric identification, as it could help us isolate other confounding factors such as temperature, which tends to follow more predictable cycles.

We use Evans and Ruf (2021) as our preferred measurement because it is available on a daily frequency, which is crucial for calculating in-utero exposure. Most other microplastic measurements that we are aware of focus on cross-sectional distribution. As discussed in Section 2, to date, in-situ measurements of microplastics are relatively limited. In Appendix Figure 1, we present a correlational analysis of the in-situ measurement compilation data provided by NOAA's NCEI Marine Microplastics product, consisting of over 8,000 measurements from various studies published since 1972, and the corresponding area's average microplastic between 2017-2018 from the Ruf data. Even in this cross-sectional correlation analysis, we observed a statistically significant correlation between the remote-sensing and in-situ measurements, for the full sample and for the post-2013 sample that were not used in training the remote-sensing measurements.

**Ocean Currents.** Our ocean currents data are from Ocean Surface Current Analyses Real-time (OSCAR) product by NASA. This dataset provides daily information on current direction and speed at a resolution of 0.25 degrees. The current dynamics are generated by combining remotely sensed signals, including sea surface height, surface vector wind, and sea surface temperature, with quasi-linear and steady flow momentum equations.

**Coastal Ocean Chlorophyll.** Our analysis includes control variables for ocean pollution. We use chlorophyll-a concentration to measure nutrient pollution, as chlorophyll presence in water can indicate levels of algae growth within a water body.[1] Chlorophyll concentration data are sourced from the Visible Infrared Imaging Radiometer Suite (VIIRS), which provides daily observations with a spatial resolution of 4km, measured in mg per m3. This satellite network monitors water bodies globally, including inland and ocean waters. We generate monthly nutrient pollution levels for coastal locations within a 100km ocean buffer.

---

[1] Chlorophyll-a can also be interpreted as a proxy for primary productivity in the ocean. Higher primary productivity often supports larger populations of zooplankton, which in turn support higher fish populations.

**Atmospheric Conditions.** To test for aerosolization of microplastics, we use aerosol optical depth (AOD) data from MERRA2 reanalysis product. This product is based on bias-corrected AOD retrieved from AVHRR, MODIS, and MISR satellite sensors, along with ground observations of AERONET AOD data. The data is at a spatial resolution of 0.5° × 0.625°.

We use evaporation data from the NCEP/DOE (National Center for Atmospheric Research/Department of Energy) Reanalysis II product. This data measures the strength of water evaporation into the atmosphere, which captures the aerosolization process from marine microplastics. The data are at the monthly level with a spatial resolution of 2.5°.

We obtain meteorological condition data from the European Centre for Medium-Range Weather Forecasts Reanalysis 5 (ERA5) products. This dataset provides hourly information on temperature and precipitation, with a grid size of 0.25°. We aggregate the data to adm1-month level.

**Seafood Access.** Seafood consumption data in the US is sourced from the Nielsen consumer panel, which consists of a representative panel of households continuously providing information about their purchases in a longitudinal study. Nielsen panelists use in-home scanners to record all their purchases. Consumers provide details about their households, the products they buy, and when and where they make purchases. Households are recorded with county, zipcode, race, and household size. We aggregate household-level purchase data into the county-month level. The primary products of interest are seafood-related items.

Fishery effort data is obtained from Global Fishing Watch. We use fishing effort measures at the vessel-grid-day level. The data product uses signals from vessel automatic identification systems (AIS). Rousseau et al. (2024) generate hours and fishing hours using national-level fishing capacity data and catch-based methods for cross-validation and integration. For each DHS cluster, US county, Mexico locality, and Brazil municipality, we calculate fishing hours within a 100km radius to measure the importance of fishery activities in the local community.

Appendix Figure 2 maps out the seafood access measurements.

**Birth.** We use microdata on birth records from four sources. First, we use data from the Demographic and Health Survey (DHS). After being matched with microplastic measurements, our DHS samples include birth records from 12 coastal countries obtained through 14 surveys. We further restrict our analysis to households that live within 200 km of the nearest shoreline, ending up with a total of 10,545 births. Each interviewee's record includes her birth month, children's birth months, health conditions, and household characteristics. Our birth outcome of interest is the existence of low birth weight, determined by raw birth weight to create a dummy variable. In the DHS responses, household locations are recorded as coordinates representing centroids of 10 km clusters. Households within a 10 km radius of each centroid share identical coordinates. Figure 2(a) visualizes the locations of DHS clusters in our sample, primarily situated in Southeast Asia and Africa, with notable variations in coverage within each country.

We obtain birth outcomes in the US from the National Vital Statistics System (NVSS). The NVSS sample includes 934,081 births from 461 coastal counties. We aggregate individual-level birth records at the county-month level for the years 2017-2018. Our primary outcome of interest is low birth weight, defined as the proportion of low birth weight births divided by the total number of new births in that county-month within the NVSS sample.

We use Mexico's birth outcome data from the Mexican National Health System, which provides microdata about birth weights, infant mortality, and mothers' demographic characteristics. The system covers the universe of births across the country and reports the locality of the mother's residence, similar to a county in the US. The data contains 725,726 birth records from 1,221 coastal municipalities.

Birth records in Brazil are sourced from the Brazil Live Birth Information System (SINASC), which provides microdata for Brazilian live births by sex, birthplace, birth weight, age, and residence of the mother. The original data is collected by the Secretariat of Health Surveillance, Ministry of Health. The Ministry requires a nationally standardized document, the "declaration of live birth" (DN), for all live births, whether the delivery occurs at home or in a hospital. Locations are recorded as municipality. The data contains 1,344,769 birth records from 2,419 coastal municipalities.

## 3.2 Econometrics

Our goal is to estimate the causal impact of microplastic exposure while in utero on birth outcome. The workhorse econometric equation is as follows:

$$\text{Low Birth Weight}_{i,t} = \beta \cdot \text{Log} \sum_{k \in [-1, -9]} \text{Microplastics}_{i,t+k}$$
$$+ \alpha_{adm1} + \alpha_{country-month} + X_{i,t}\gamma + \varepsilon_{i,t} \quad (1)$$

where Low Birth Weight$_{i,t}$ is an indicator variable for whether birth i occurring at month t is weighing less than 2,500 grams. $\sum_{k \in [-1, -9]} \text{Microplastics}_{i,t+k}$ captures i's exposure to microplastics while in utero, which equals the logged sum of microplastic levels observed at the nearest coastal sea grid to i's birth location over the 9-month period preceding the birth. For each birth, we match it to the 0.25 degree grid in the microplastic data that are the closest to the birth's location. Birth locations are measured by survey cluster latitude and longitude in the DHS data, county centroid in the USA data, municipality centroid in the Brazil data, and municipality centroid in the Mexico data.

$\alpha_{adm1}$ are region fixed effects. We define regions as "adm1" administrative level-1 subregion (e.g., state for the U.S. and Estados for Brazil and Mexico). $\alpha_{country-month}$ are country by month-of-sample fixed effects that control for country-specific seasonal and secular trends. $X_{i,t}$ includes time-varying environmental covariates, which helps rule out confounding effects from other factors that might correlate with microplastic exposure while might also having independent impacts on birth outcomes, including temperature, precipitation, air pollution, and ocean chlorophyll-a. $\varepsilon_{i,t}$ is the error term. Standard errors are clustered at the admin1 level.

The key coefficient of interest is thus $\beta$, which captures the impact of in-utero exposure to microplastics on likelihood of low birth weight. The challenge in interpreting $\beta$ as the causal effect of microplastics lies in potential endogeneity: some third factors correlated with microplastics are directly affecting birth outcomes and are not captured by the fixed effects and co-pollutant controls. This concern is less likely to be substantial compared with other forms of environmental pollution because it takes at least several decades for microplastics to form. As a result, the third factors affecting microplastic generation may be correlated with socioeconomic conditions decades ago, but not recent ones, and are less likely to directly impact birth outcomes. Moreover, marine microplastics are often generated elsewhere, so the factors that produce microplastics tend to be distant from the areas of interest with birth outcomes. We address these endogeneity concerns in several ways.

<u>First</u>, we construct alternative measures of in-utero microplastic exposure to leverage the quasi-random variation that comes from ocean dynamics. Specifically, we build an ocean current dynamic model that computes predicted coastal microplastic concentrations resulting from transported microplastics from distant areas. For example, we start with observed microplastic concentrations in areas that are at least 200 km away from coastal zones and use ocean currents data to simulate the extent to which these far-sea microplastics may travel toward coastal areas. Using this "transported" microplastic variation in regression equation (1) alleviates endogeneity concerns because the source of variation—microplastic variation far from land and changes in ocean currents—should have little to do with coastal birth weights except through their influence on coastal microplastic variability. We will describe the construction of transported microplastics in Section 3.3 that follows.

Second, we estimate an augmented version of equation (1) by replacing the in-utero, 9-month microplastic exposure with separate trimester exposure terms. These trimester terms provide additional information on which trimester exposure seems to matter the most, in a way that we can compare to the medical literature. For example, we know that fetal weight is mostly determined toward the later periods of pregnancy, and therefore, we would expect a smaller effect, if any, during the first trimester. We further add a "preconception" and a "postpartum" term, measuring microplastic "exposure" in the 3 months prior to conception and 3 months after birth. These "placebo" effect terms provide a chance to test for model misspecification. The augmented estimation equation is as follows:

$$
\begin{aligned}
\text{Low Birth Weight}_{i,t} = \ & \beta^{\text{Preconception}} \cdot \text{Log} \sum_{k \in [-10,-12]} \text{Microplastics}_{i,t+k} \\
& + \beta^{\text{1st Trimester}} \cdot \text{Log} \sum_{k \in [-7,-9]} \text{Microplastics}_{i,t+k} \\
& + \beta^{\text{2nd Trimester}} \cdot \text{Log} \sum_{k \in [-4,-6]} \text{Microplastics}_{i,t+k} \\
& + \beta^{\text{3rd Trimester}} \cdot \text{Log} \sum_{k \in [-1,-3]} \text{Microplastics}_{i,t+k} \\
& + \beta^{\text{Postpartum}} \cdot \text{Log} \sum_{k \in [0,2]} \text{Microplastics}_{i,t+k} \\
& + \alpha_{\text{adm1}} + \alpha_{\text{country-month}} + X_{i,t}\gamma + \varepsilon_{i,t} \quad (2)
\end{aligned}
$$

where the coefficients of interest are $\beta^{\text{Preconception}}$, $\beta^{\text{1st Trimester}}$, $\beta^{\text{2nd Trimester}}$, $\beta^{\text{3rd Trimester}}$, and $\beta^{\text{Postpartum}}$.

Third, our estimation sample contains data from multiple countries spanning different parts of the world. This means the characteristics of microplastic fluctuations vary significantly across locations in our sample (Figure 2), as do other environmental and economic conditions. Most importantly, what constitutes omitted factors in one area is unlikely to prevail in other countries. To the extent that we observe robust effect estimates in distinct parts of the world, it enhances our confidence that our model is capturing the genuine impact of microplastics.

## 3.3 Microplastic Transport Modeling

To capture the microplastics transported from the far sea to coastal areas, we apply an oceanography model to construct an ocean current flow intensity matrix. Here we provide an intuitive explanation of the procedure, leaving computational details to the Technical Appendix.

The input data is obtained from the Ocean Surface Current Analysis Real-time (OSCAR) product. This dataset provides daily information on current direction and speed at a resolution of 0.25-degree. Using this input, we first generate a spatial representation of current *flows* from individual current *vectors*. Beginning from a particular grid and day, we construct streamlines by sequentially following the current speed and direction on a daily basis. This process maps out the evolving trajectories of the ocean current field, giving us daily representations of the distribution of current flow intensity across the world.

Our goal is to build an "exogenous" measure of microplastic exposure that is transported from far sea areas (and thus less likely to be correlated with activities in the coastal areas of interest, which we worry might have independent impacts on birth outcomes and confound our estimation of the effect of microplastics.) This variable, $MP_{i \to r, m}$, is a summary of downstream intensity blowing from a sender grid $i$ to a receiver coastal grid $r$ in month $m$. Note that, for any pair of grids that are significantly distant from each other, it is meaningless to talk about up/downstream relationship on any given "day" because ocean currents moving from the sender grid may take days to arrive at the receiver grid. We therefore track the trajectory of currents "originating" from a sender grid and their impacts of downstream grids over multiple days – or "steps" as we refer to them below – using the ocean current streamline data that we constructed earlier.

To define far sea senders, we use grids that are at least 200 km to the shoreline. In an alternative specification, we use all grids as potential contributors. The standard bias-variance tradeoff prevails here: restricting to farther away contributor grids reduces predictability of far-sea microplastic condition on coastal condition, but increases exogeneity. After the sender list is specified, we define downstream intensity score between each sender and each coastal receiver using the following formula:

$$\text{Current}_{i \rightarrow r,d,t} = \exp\{-\alpha \cdot \text{rad}_t - \beta \cdot |\theta|_{i \rightarrow r,d,t} - \gamma \cdot \text{dist}_{i \rightarrow r,d,t}\} \quad (3)$$

Starting from a sender grid $i$ on day $d$ and at step $t$, we assume downstream intensity of receiver grid $r$ follows an exponential decay as a function of three components (U.S. EPA, 2018; Phillips et al., 2021). The first component is the search radius at the step ($\text{rad}_t$), which captures general decreases of downstream intensity over steps. The initial radius is 1 degree which is about 111km at the equator, and we increase the search radius by 0.05 degree (about 6km at the equator) at each step to capture both the uncertainty in the streamline computation and the dispersion of microplastics in the ocean. The second component is the scalar product of the angle between the receiver grid and the ocean current direction originating from the sender's location ($|\theta|_{i \rightarrow r,d,t}$), which means we assign higher intensity to receiver grids that sit closer to the exactly-downstream direction of the sender grid. The third component is simply the distance between the sender and the receiver grid ($\text{dist}_{i \rightarrow r,d,t}$), which captures geographic decay. We assume $\text{Current}_{i \rightarrow r,d,t}$ to be zero if $d_{i \rightarrow r,d,t} > \text{rad}_t$ (i.e., if receiver grid lies outside of the search radius at step $t$) or if $\theta_{i \rightarrow r,d,t} > 0.4 \text{ radian}$ (i.e., if the receiver grid is not obviously in the downstream direction from the sender grid.[2] Starting from each particular sender grid and day, we iterate the procedure for 90 steps (i.e., three months).

A visualization of the procedure is shown in Figure 3. The red arrow at the center represents the locus of the current flow starting from the sender grid. The growing ball of uncertainty around the arrow shows expanding search radius $\text{rad}$ over steps in equation (3). The visualization also explains why both the relative angle and distance variables ($\theta$ and $\text{dist}$) have starting day and step subscripts ($d$ and $t$): we track where ocean currents originate and where they move to, and we compute each receiver grid's relative angle and distances dynamically.

We aggregate step-wise downstream intensity scores to the day level:

$$\text{Current}_{i \rightarrow r,d} = \sum_{d+t=d} \text{Current}_{i \rightarrow r,d,t} \quad (4)$$

---

[2] We use parameter values $\{\alpha, \beta, \gamma\} = \{0.8, 0.49, 0.23\}$. These numbers are empirically determined such that we would obtain a spatially continuous flow coefficient function through the successive steps, and that the directionality of the observed currents are respected through the flow coefficient representation. See Online Appendix for more details.

and in the econometric analysis below, we further average this city pair-daily score to the monthly frequency to make the size of the regression dataset manageable. We then calculate the imported microplastics (MP) concentrations $MP_{i \to r,m}$ by multiplying $Current_{i \to r,m}$ with $MP_{i,m}$. Then we aggregate all sender grids, and in the subsequent we will use $\sum_i MP_{i \to r,m}$ as an exogenous regressor to measure total transported microplastics received by grid $r$ in month $m$ and to estimate its impact on coastal infant health.

Let us reiterate the main idea: in some versions of the estimation of equation (1), we propose to use "transported" microplastics $\sum_i MP_{i \to r,m}$ – the portion of variation in $MP_{r,m}$ that is likely driven by exogenously transported microplastics from the far sea – as the main regressor rather than using $MP_{r,m}$ directly which we worry might be endogenously related to other determinants of coastal infant health. Our claim here is that the ocean current modeling exercise has achieved the objective of teasing out that exogenous variation. One way to see this is to examine how $MP_{i,m}$ (sender grid's microplastic concentration), $Current_{i \to r,m}$, and $MP_{r,m}$ (receiver grid's microplastic concentration) are empirically related to each other. Appendix Figure 3 reports the passthrough between $MP_{i,m}$ and $MP_{r,m}$ as a function of bins of $Current_{i \to r,m}$. We estimate the following regression equation:

$$Log\ MP_{r,m} = \beta \cdot Current_{i \to r,m} \times Log\ MP_{i,m}$$
$$+ \gamma \cdot Current_{i \to r,m} + \delta \cdot Log\ MP_{i,m} + \alpha_{ir} + \alpha_m + \varepsilon_{i,r,m} \quad (5)$$

where $Current_{i \to r,m}$ enters the regression decile bins, with the 10th bin representing the weakest ocean current condition being the reference category and thus omitted from the regression. Equation (5) includes sender-by-receiver fixed effects ($\alpha_{ir}$), capturing the identifying variation from within sender-receiver pairs across different months, where variations in ocean current intensities are plausibly exogenous. The pattern of the $\beta$'s in Appendix Figure 3 suggests that indeed the receiver grid's measured microplastic concentration respond most strongly to sender grid's microplastic concentration when the modelled oceanic transport from the sender grid to the receiver grid is stronger.

It is perhaps worth noting that, although *ex ante* we thought endogeneity of microplastic exposure may pose a significant identification issue, *ex post*, using local microplastic measure, transported microplastics from all other grids, and transported microplastics from sender grids over 200 km away all give rise to very similar estimation results. This pattern suggests the endeneity issue in our estimation context was not severe after all. We have given more thoughts on why this is in Section 3.2.

# 4. Results

## 4.1 Main Results

Figure 4 provides a first look into our main estimation results based on equation (2), linking each individual's birth outcome (i.e., whether the birth has low birth weight of below 2,500 grams at birth) to in-utero exposure to microplastic pollution. From left to right, the coefficients represent the impact of microplastic exposure over the preconception quarter, the first, second, and third trimesters, and the postpartum quarter. We are mainly interested in the effects during the *actual* pregnancy (i.e., the three trimesters). These trimester terms provide information on which trimester exposure seems to matter the most, in a way that we can compare to the medical literature. For example, we know that fetal weight is mostly determined toward the later periods of pregnancy, and therefore, we would expect a smaller effect, if any, during the first trimester.

The preconception and postpartum terms measure microplastic "exposure" in the 3 months prior to conception and 3 months after birth. These coefficients provide useful *placebo* tests for our model specification: we do *not* expect microplastic concentration during these periods to matter for birth outcomes, unless there are model misspecifications that lead to spurious findings. Results in Figure 4 suggest that the impact on low birth weight is most attributable to microplastic exposure during the second and third trimesters, consistent with the notion that fetal development is mostly determined in these periods. Reassuringly, we do not find significant effects from the two placebo periods.

Table 1 reports more details on the microplastic-health links. Each coefficient reported in this table represents an estimate of $\beta$ from equation (1) with varying microplastic measurement, sample restrictions, and control variables. Start with column 1. Here we report estimation results using pooled data. Each coefficient corresponds to a separate regression using a different right-hand-side microplastic measurement, as indicated by the row names. The number on the top row shows that a log increase in in-utero microplastic exposure, measured directly using observed local variation at the birth's coastal location, leads to an increase in the likelihood of low birth weight by 0.287 per 1,000 births. The number on the middle row shows the estimate when we use transported microplastics as the independent variable, where the effect size is a 0.442 per 1,000 births increase in low birth weight per log increase in microplastics. The bottom row also uses transported microplastics as the independent variable but restricts contributing grids to those that are at least 200 km away from the coastal location. Here the effect size is 0.381. The bottom section of column 1 shows that this regression does not include additional environmental controls (temperature, precipitation, atmospheric aerosol, and coastal ocean chlorophyll). The mean of the dependent variable is 27.6, meaning on average the rate of low birth weight in the pooled data is 27.6 per 1,000 births. The average effect sizes from these three regressions, (0.287+0.442+0.381)/3=0.370, therefore corresponds to about a 2% increase in the odds of low birth weight for a log increase in in-utero microplastic exposure. The bottom row shows that the regressions include a little over 3 million individual births.

The rest of the columns are variants of the baseline in column 1. The estimation in column 2 is identical to column 1 except that it includes additional environmental controls. Columns 3-4, 5-6, 7-8, and 9-10 report subsample estimation results for the DHS samples, the USA sample, the Brazil sample, and the Mexico sample, respectively. We find statistically significant effects for the majority of the samples and econometric specifications. On the "percent change in likelihood of low birth weight per 1 log increase in in-utero microplastic exposure" scale, the effect sizes are 1.2 percent for the USA, 1.9 percent for Brazil, 2.1 percent for Mexico, and 8.1 percent for DHS countries. We are generally underpowered with the DHS data, which features a much smaller sample size (about 10,500), and so the noisy estimates may have exaggerated the true effect size, but they are of a similar order of magnitude to the other samples. Scaled linearly, our estimate implies that about 205,800 of LBW coastal births annually worldwide can be attributed to microplastic exposure.

Our estimates using the local versus transported microplastic measures yield similar results, suggesting that the degree of endogeneity—such as omitted variable biases—may not pose a significant problem in equation (1). This is also supported by the fact that our estimates are largely unchanged after we condition on a rich set of environmental covariates.

## 4.2 Channels

How does in-utero exposure to microplastics cause low birth weight? In this section, we introduce additional data to test two prevailing hypotheses about the health impact channels of microplastics.

**Access to Seafood.** The first obvious channel is seafood consumption. Microplastics deposit in sea animals and accumulate as they move up the food chain, ultimately entering the human system as seafood is consumed. One intuitive test of this channel is to see whether the impact of microplastics on low birth weight is higher in areas with a higher rate of seafood consumption. We come up with two such measures. The first measure is derived from Nielsen retail scanner data, which we use to construct county-level seafood spending per capita. This data only exists for the USA. The second measure comes from Global Fishing Watch, where we use an area's 2017-2018 average fishing hours as a proxy for access to seafood. This data is available for all locations. In Table 2, we repeat the estimation in column 1 of Table 1, but interact the microplastic measures with these two seafood consumption proxies separately. We find no evidence of substantial heterogeneity in the health impacts of microplastics across areas with high versus low seafood consumption.

In a similar vein, one other channel lies in seafood trade, as seafood consumed by US consumers is mainly imported from other countries. As a result, US health outcomes may be affected by microplastics in exporting countries rather than local pollution. To test the trade channel, we use seafood importing data from the USA Trade Online, provided by the US Census Bureau. The variable of interest is import value at the US state-exporting country-month level. For each birth record in state $i$ in month $m$, we merge it with seafood exporters' microplastics, a weighted average of microplastics near each exporting country $j$ in month $m$. This treatment measure is calculated using three steps: i) We retrieve microplastics near each exporting country-month using a 200km buffer around country shorelines, i.e. $MP_{j,m}$; ii) Weight is calculated as importing value of live seafood, HS code under 0301 and 0302, at the state-country-month level, i.e. $ImSeafood_{i,j,m}$; iii) To aggregate over exporting countries, we calculate the weighted average: $\sum_j (ImpSeafood_{i,j,m} * MP_{j,m}) / \sum_j ImpSeafood_{i,j,m}$.

In Appendix Table 1, we put both self area microplastics and seafood exporters' microplastics on the right-hand side to run a horse-race. Estimates on local microplastics are large, precise, and of similar magnitude to those in Table 1 column (5) and (6), confirming the robustness of our observed pattern between local microplastics and local low birth weight incidences. Estimates on exporters' microplastics are imprecise and close zero, which suggests little evidence that exporter's surrounding microplastics affects consumption states' birth outcomes.

Heterogeneity tests and the lack of effects of exporters' microplastics here do not necessarily mean that exposure has no health impacts through seafood consumption. For example, newborns who are exposed to microplastics may exhibit other adverse effects later in life, even in the absence of a birth weight effect. Another explanation is endogenous fishing efforts or product quality control, which may mitigate the adverse effects of microplastics on seafood quantity and quality and result in little effects on human health. More scientific and econometric evidence is needed on the possibility of mother-to-child transmission of microplastic intake, and how microplastic deposits in the fetus may impact short- and long-term outcomes.

**Aerosolization.** A more novel channel we test next is microplastic aerosolization. Due to their small sizes, many microplastic species may become airborne under suitable atmospheric conditions. Aerosolized microplastics may thus become inhalable and enter the human system through respiratory tracts and, like fine particulate matter, penetrate human organs. This could be a potentially important channel, as prior research has established the causal link between airborne particle pollution and adverse infant health outcomes (e.g., Jayachandran, 2009; Currie and Schwandt, 2016; Alexander and Schwandt, 2022).

To test aerosolization, we obtain a monthly remote-sensing measure of aerosol optical depth at coastal locations included in our study sample. We then test whether higher coastal microplastic concentration is correlated with higher aerosol pollution. To tease out causality, in alternative specifications we further use transported microplastics from the far sea (as we did in the health regressions) to predict coastal aerosol pollution. Because vertical air motion is likely a key driver of aerosolization, in some of our estimations we also interact microplastic concentration with the rate of evaporation observed at coastal locations (which is largely an increasing function of temperature and plausibly random conditional local seasonality). Table 3 reports estimates that support aerosolization: higher microplastic predicts coastal aerosol pollution (columns 1, 3, and 5), and this effect is stronger when higher rates of evaporation are observed (columns 2, 4, and 6). These findings point to the importance of further understanding the medical underpinnings of the impact of plastic particles and fibers on respiratory and other health outcomes.

The interactive effect between sea surface evaporation and microplastic aerosolization also implies that population exposure to airborne microplastic pollution is expected to rise as climate change continues to cause a warm up of ocean temperature and evaporation (Abraham et al., 2013; Trenberth, 2011).

# 5. Conclusion

Decades of academic and policy debate have been paid to the issue of plastic pollution but there is a paucity of direct evidence on its health effects. Growing availability of high-resolution measurements of microplastics, oceanic motion, and health, as well as methods that are able to identify causal signals in these data, may help close this gap. Our initial look into this question in infant health setting suggests microplastics exposure may indeed have important health ramifications as science from multiple fields has warned.

**Figure 1. Plastic Pollution Sources, Ambient Presence, and Exposure**

(a) Plastic pollution sources (<u>European Environment Agency, 2022</u>)

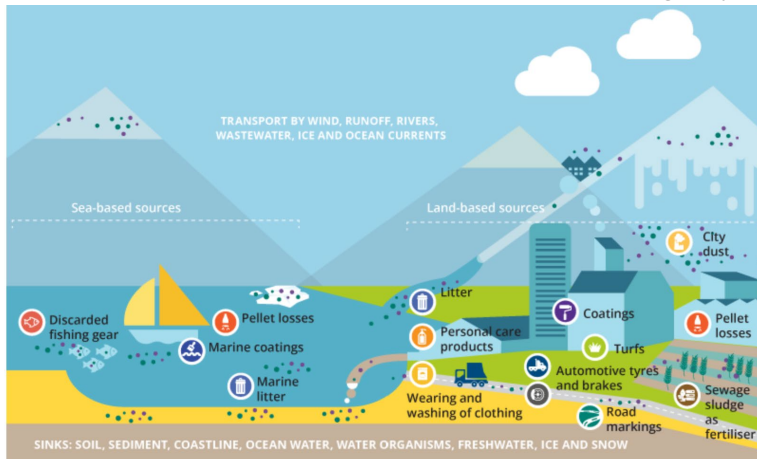

(b) Microplastic ingestion in fish larvae (<u>Steer et al., 2017</u>)

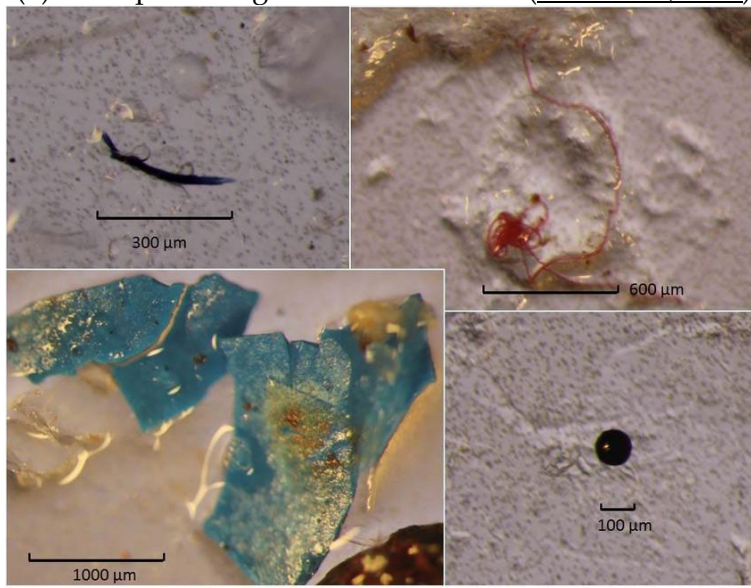

(c) Airborne microplastic (<u>Vianello et al., 2019</u>)

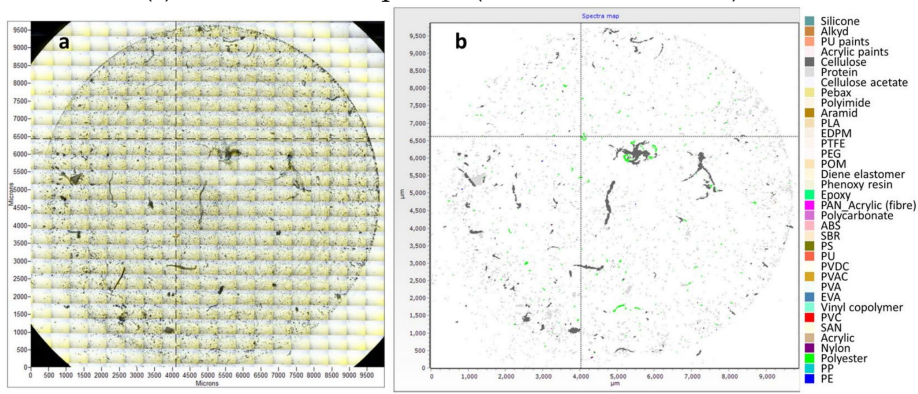

*Notes:* Panel (a) shows sources of plastic pollution emissions. Panel (b) shows categorization of plastic particles and potential marine animal exposure. Panel (c) shows a microscopic illustration of airborne microplastic particles.

**Figure 2. Remote-Sensing Microplastic Measurement: Summary**

(a) Global distribution

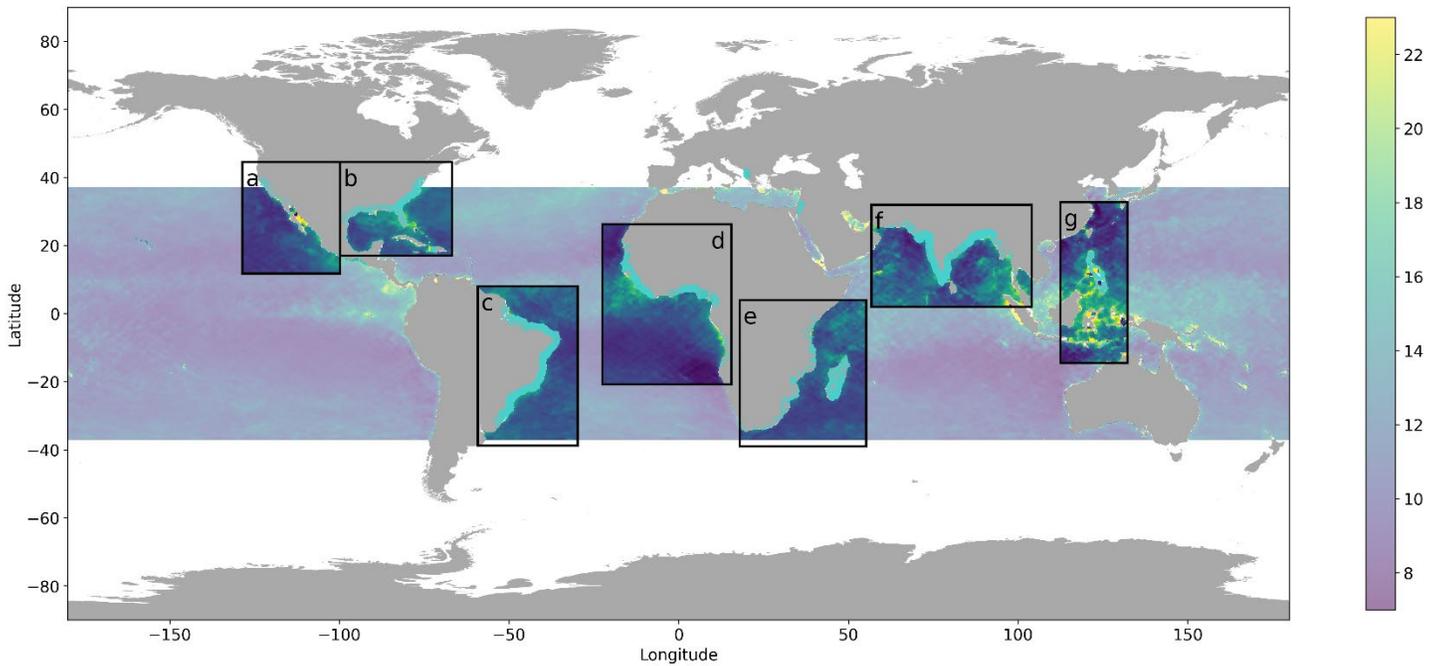

(b) Time trends in study areas

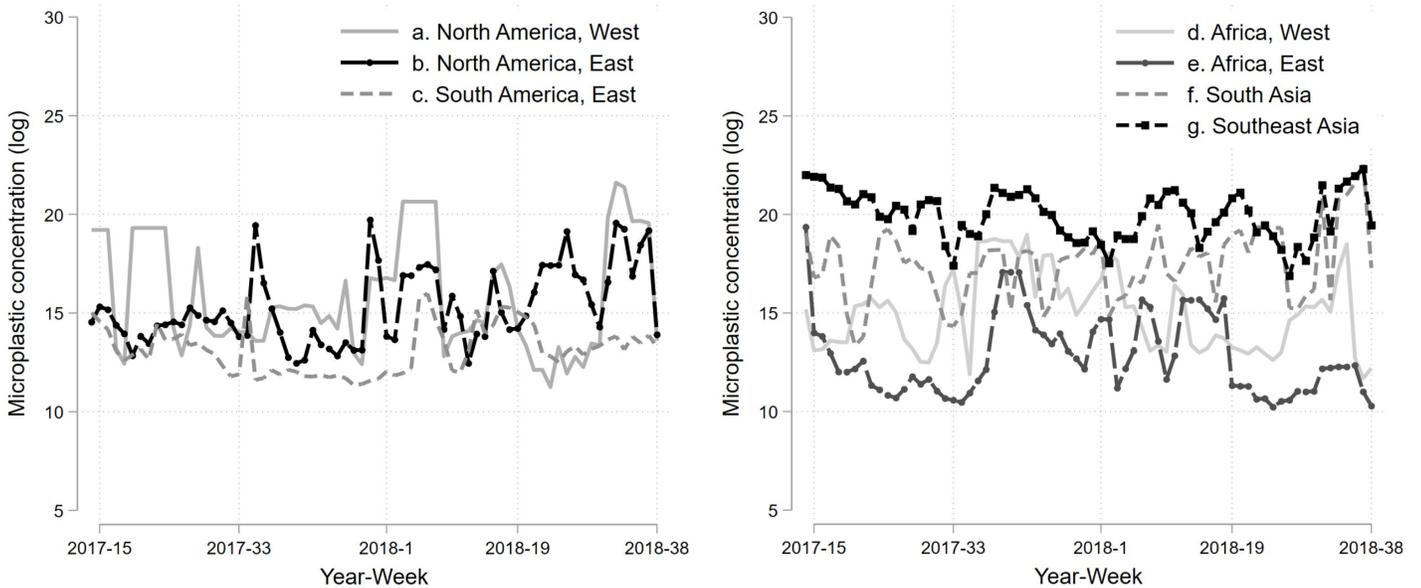

*Notes:* Panel (a) shows geographic distribution of microplastic concentration from the remote-sensing measurement. Panel (b) reports average microplastic concentration trends in seven selected regions where our study samples concentrate.

## Figure 3. Transported Microplastic Modeling

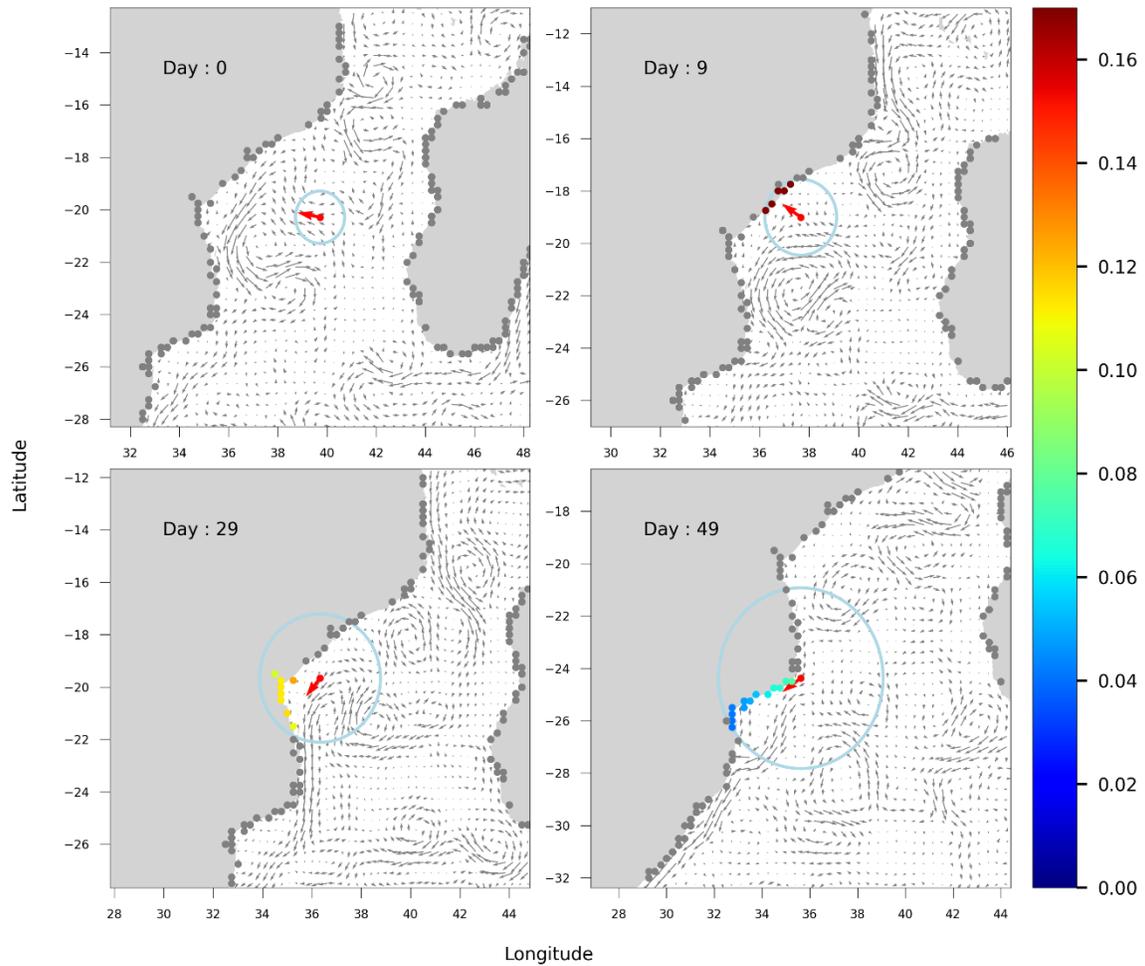

*Notes:* This figure provides an illustration of the transport modeling that tracks coastal variation in microplastic exposure due to microplastic presence from sea surface afar. Shaded gray areas represent land. Dots represent coastal locations. Panels track a source grid and show its location (red arrow), area of influence (blue circle), and coefficient-of-influences (colored dots) at four different time steps.

**Figure 4. The Effect of In-Utero Microplastic Exposure on Low Birth Weight**

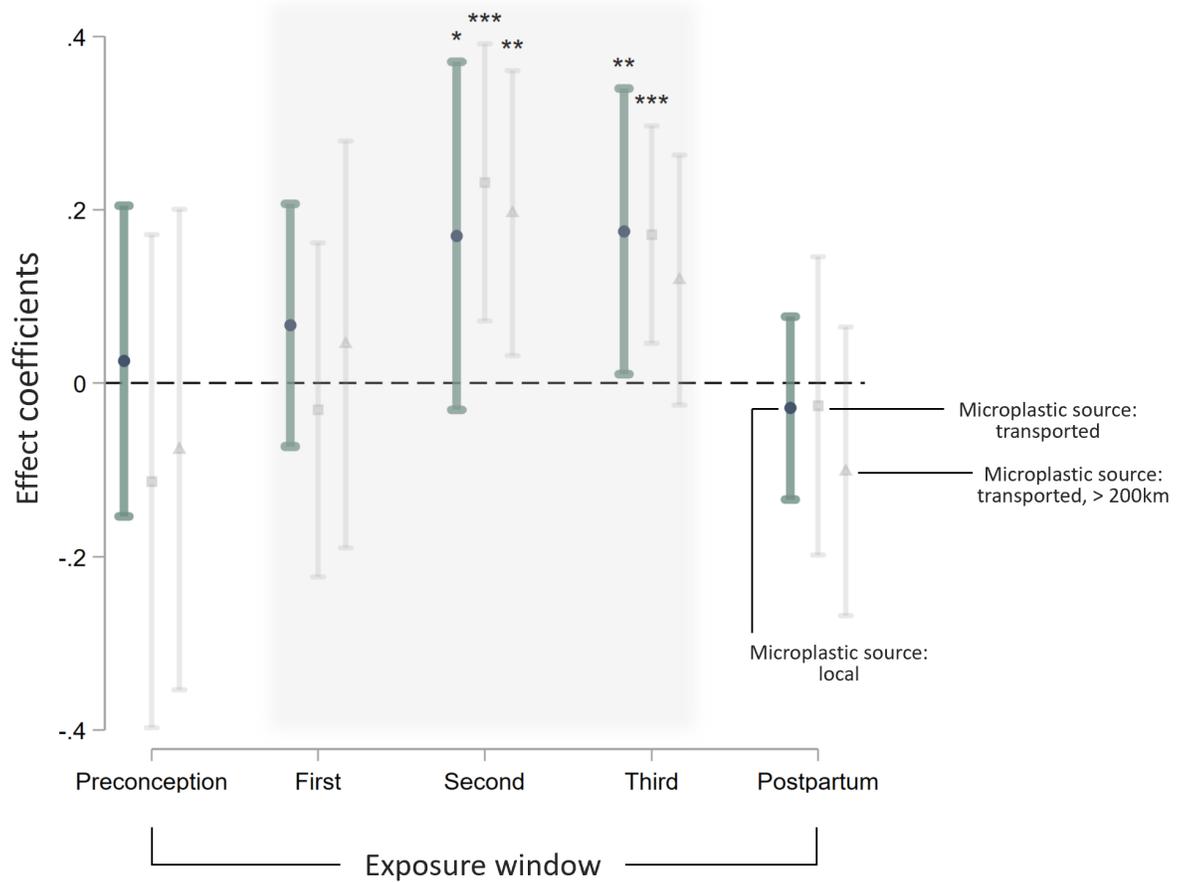

*Notes:* This figure shows coefficient estimates from three separate regressions looking at the impact of in-utero microplastic exposure on incidents of low birth weight. The regressions use different microplastic measurement (circles for local microplastic source, squares for transported source, and triangles for transported source with a 200km buffer) but are otherwise with identical specification. "Preconception" corresponds to microplastic exposure during the 3-month period prior to conception, and "Postpartum" corresponds to exposure during the 3-month periods after birth. *: $p < 0.10$; **: $p < 0.05$; ***: $p < 0.01$.

**Table 1. The Effect of In-Utero Exposure to Microplastics on Low Birth Weight**

| | (1) | (2) | (3) | (4) | (5) | (6) | (7) | (8) | (9) | (10) |
|---|---|---|---|---|---|---|---|---|---|---|
| | Pooled data | | DHS sample | | USA sample | | Brazil sample | | Mexico sample | |
| Log microplastics (local) | 0.287** | 0.321*** | 5.813* | 4.210 | 0.314* | 0.582*** | 0.306 | 0.520** | 0.084 | 0.215 |
| | (0.111) | (0.111) | (3.418) | (2.769) | (0.156) | (0.122) | (0.215) | (0.216) | (0.350) | (0.480) |
| Log microplastics (transported) | 0.442*** | 0.443*** | 5.511 | 0.334 | 0.382*** | 0.376*** | 0.588 | 0.503 | 0.462*** | 0.605*** |
| | (0.078) | (0.072) | (8.401) | (6.975) | (0.078) | (0.064) | (0.352) | (0.395) | (0.113) | (0.098) |
| Log microplastics (transported from > 200km) | 0.381*** | 0.373*** | 11.291 | 6.116 | 0.252*** | 0.230*** | 0.780*** | 0.732** | 0.367*** | 0.520*** |
| | (0.077) | (0.076) | (7.327) | (6.969) | (0.059) | (0.062) | (0.248) | (0.272) | (0.117) | (0.126) |
| Co-pollutants controls | | ✓ | | ✓ | | ✓ | | ✓ | | ✓ |
| FEs: admin 1 region | ✓ | ✓ | ✓ | ✓ | ✓ | ✓ | ✓ | ✓ | ✓ | ✓ |
| FEs: country × month of sample | ✓ | ✓ | ✓ | ✓ | ✓ | ✓ | ✓ | ✓ | ✓ | ✓ |
| Dep. var. mean | 27.6 | 28.0 | 68.6 | 68.0 | 30.6 | 30.6 | 30.2 | 30.2 | 18.2 | 17.6 |
| Observations | 3,015,121 | 2,830,110 | 10,545 | 10,320 | 934,081 | 934,081 | 1,344,769 | 1,332,852 | 725,726 | 552,857 |

*Notes:* Each cell is a separate regression. In each regression, the dependent variable is a dummy for low birth weight. Independent variable is a type of microplastic exposure measurement (local, transported, or transported from grids over 200km away). All regressions control for GADM admin level-1 subcountry region fixed effects and country by month-of-sample fixed effects. "Co-pollutants controls" include logged average temperature, precipitation, aerosol pollution, and coastal chlorophyll levels over the course of pregnancy. Standard errors are clustered at the admin level-1 level. *: $p < 0.10$; **: $p < 0.05$; ***: $p < 0.01$.

## Table 2. Channel: Access to Seafood

|  | (1) | (2) | (3) | (4) | (5) | (6) |
|---|---|---|---|---|---|---|
| Log microplastics | 0.283 (0.189) | 0.275** (0.111) | 0.389*** (0.090) | 0.355*** (0.116) | 0.273*** (0.071) | 0.236* (0.135) |
| Log microplastics × Log seafood spending | -0.102 (0.175) |  | -0.025 (0.068) |  | 0.100 (0.070) |  |
| Log microplastics × Log fishing hours |  | -0.006 (0.083) |  | 0.014 (0.056) |  | 0.006 (0.045) |
| Microplastics source | local | local | transported, all grids | transported, all grids | transported, >200km | transported, >200km |
| Observations | 914,770 | 1,944,369 | 914,114 | 1,977,857 | 914,114 | 1,977,857 |

*Notes:* Each column is a separate regression. In each regression, the dependent variable is a dummy for low birth weight. "Log seafood spending" is logged US county 2017-2018 average seafood consumption per capita measured from Nielsen scanner data (columns 1, 3, and 5). "Log fishing hours" is logged 2017-2018 average fishing hours measured from the Global Fishing Watch automatic identification system data (columns 2, 4, and 6). All regressions control for GADM admin level-1 subcountry region fixed effects and country by month-of-sample fixed effects. Standard errors are clustered at the admin level-1 level. *: $p < 0.10$; **: $p < 0.05$; ***: $p < 0.01$.

## Table 3. Channel: Microplastic Aerosolization

|  | (1) | (2) | (3) | (4) | (5) | (6) |
|---|---|---|---|---|---|---|
| Log microplastics | 0.0217*** (0.0006) | 0.0143*** (0.0007) | 0.0451*** (0.0010) | 0.0371*** (0.0011) | 0.0435*** (0.0009) | 0.0358*** (0.0011) |
| Log microplastics × Log evaporation |  | 0.0136*** (0.0009) |  | 0.0113*** (0.0016) |  | 0.0071*** (0.0014) |
| Microplastics source | local | local | transported, all grids | transported, all grids | transported, >200km | transported, >200km |
| Observations | 143,042 | 83,808 | 142,608 | 83,650 | 141,773 | 83,297 |

*Notes:* Each column is a separate regression. In each regression, the dependent variable is logged aerosol optical death measured at the coastal grid. "Log evaporation" is logged rate of evaporation measured at the coastal grid. All regressions control for grid fixed effects and country by month-of-sample fixed effects. Standard errors are clustered at the grid level. *: $p < 0.10$; **: $p < 0.05$; ***: $p < 0.01$.

**Appendix Figures and Tables**

## Appendix Figure 1. Remote-Sensing Microplastic Measurement: Validation

### (a) In-situ microplastic sampling locations

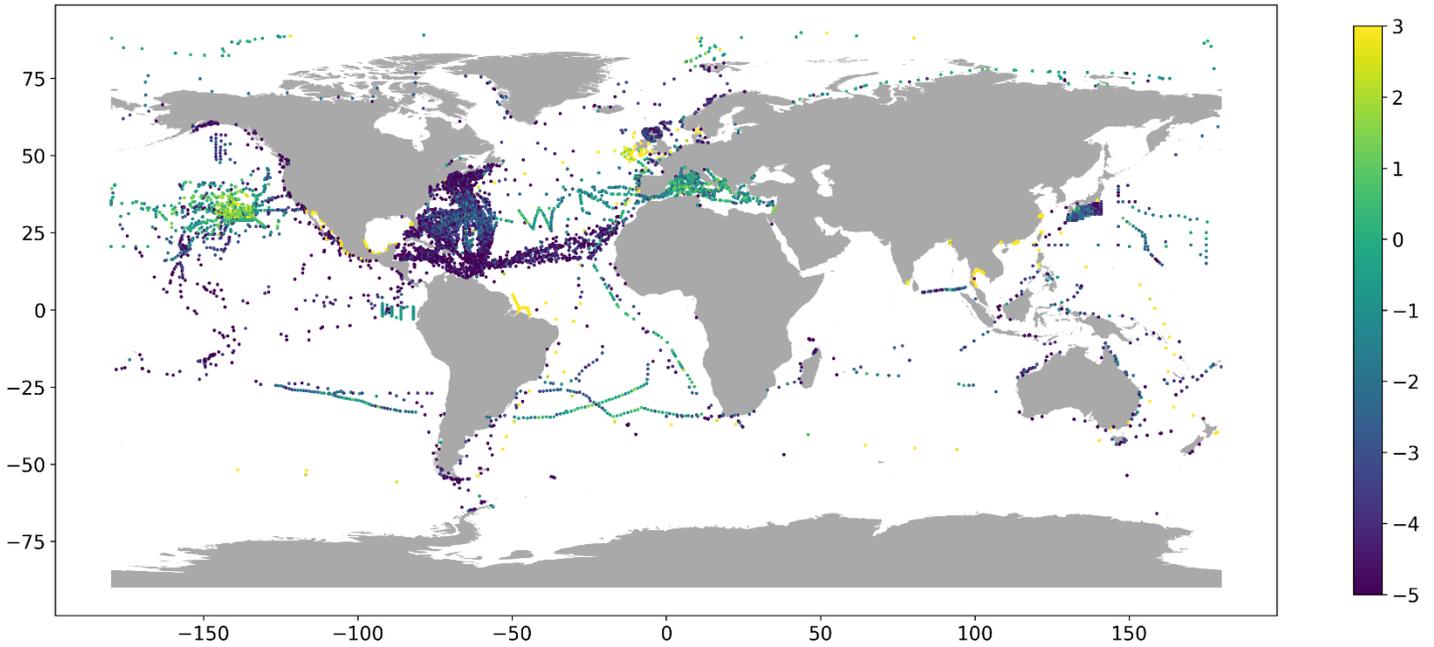

### (b) Correlation between in-situ and remote-sensing measurements

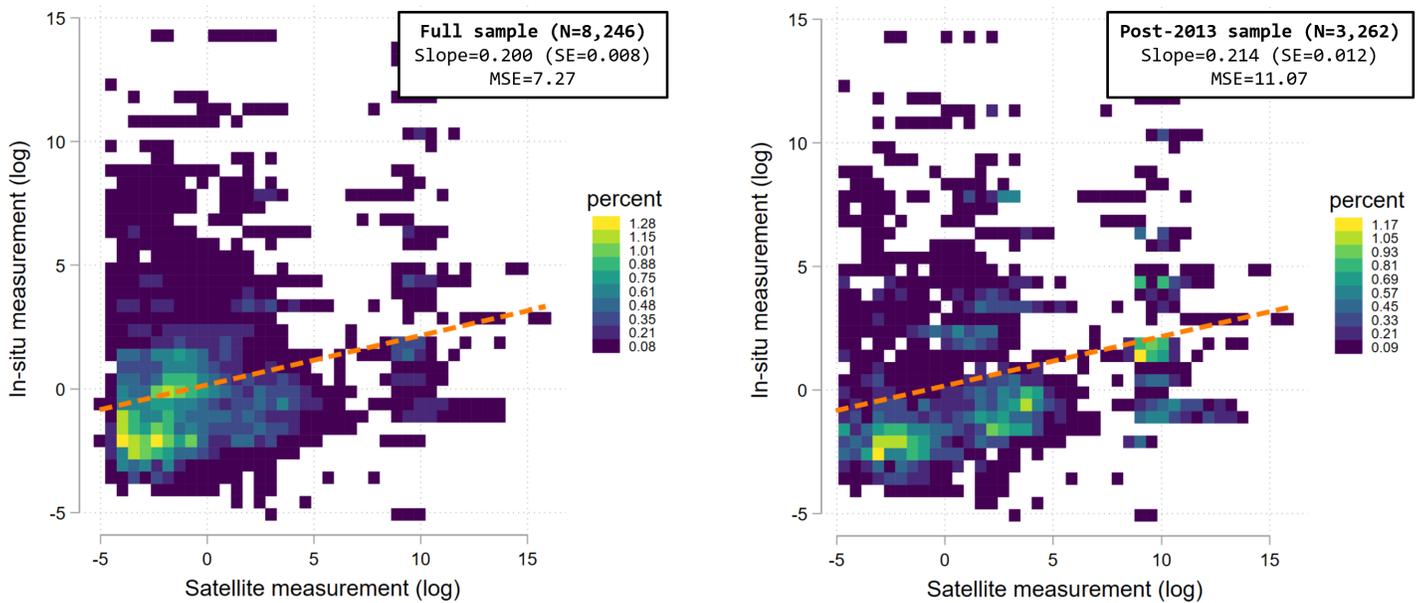

*Notes:* Panel (a) shows geographic distribution of in-situ microplastic sampling locations and recorded concentrations. Panel (b) reports correlation between the in-situ and remote-sensing measurements at collocated areas. Left panel shows full-sample results. Right panel restricts to post-2013 periods where the in-situ measurements are not used in training the remote-sensing measurement.

## Appendix Figure 2. Seafood Access Measurements

### (a) Per capita seafood spending, 2017-2018

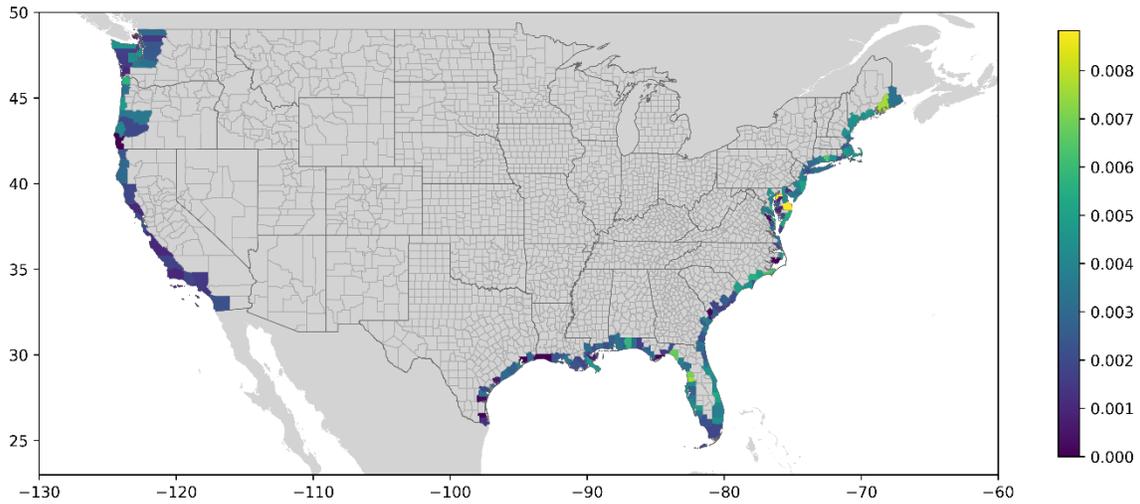

### (b) Average fishing hours within 100 km of the coastal location, 2017-2018

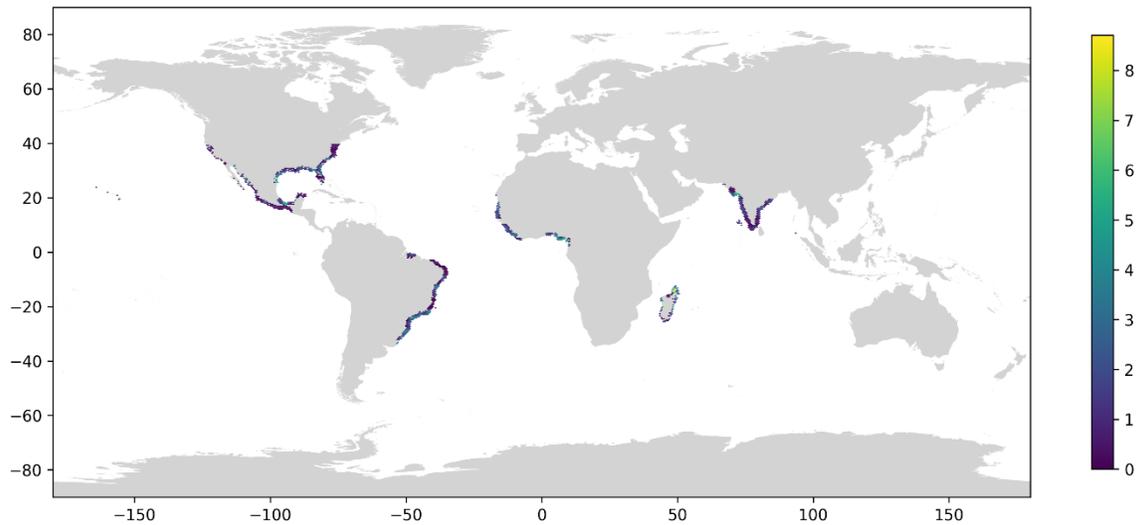

*Notes:* Panel (a) shows county level 2017-2018 average per capita seafood spending from Nielsen scanner data. Panel (b) shows 2017-2018 average fishing hours from Global Fishing Watch Automatic Identification Systems (AIS) data.

**Appendix Figure 3. Passthrough of Microplastic from Ocean to Coastal Locations by Downstream Score Deciles**

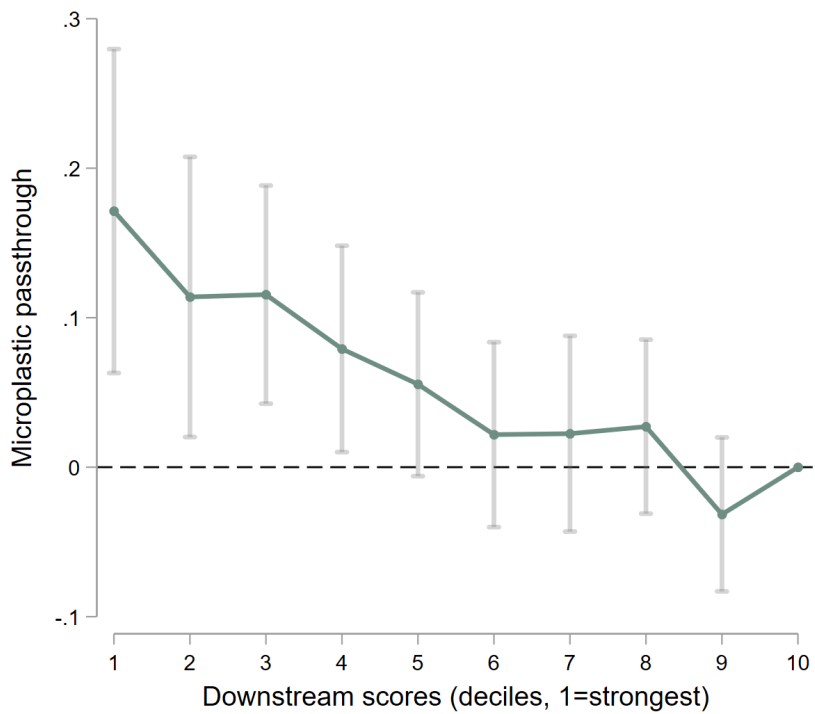

*Notes:* This figure shows coefficients from a regression of a coastal location's log microplastic concentration on an upstream ocean grid's log microplastic concentration, with the effect allowed to vary by the downstream score from the upstream location to the coastal location according to our ocean current model. All regression controls for location pair and month-of-sample fixed effects. Standard errors are three-way clustered at the sender grid, coastal grid, and month-of-sample levels. Range bars show 95 percent confidence intervals.

**Appendix Table 1. Impacts of Local versus Seafood Exporter Countries' Microplastics on Low Birth Weight (USA Sample)**

|  | (1) | (2) | (3) |
|---|---|---|---|
| Log self area microplastics | 0.646*** | 0.407*** | 0.266*** |
|  | (0.058) | (0.086) | (0.072) |
| Log seafood exporters' microplastics | -0.048 | 0.063 | 0.022 |
|  | (0.117) | (0.125) | (0.133) |
|  |  |  |  |
| Self-area microplastics source | local | transported, all grids | transported, >200km |
| Observations | 552,706 | 552,050 | 552,050 |

*Notes:* Each column is a separate regression. In each regression, the dependent variable is a dummy for low birth weight. Independent variable "Log self area microplastics" is a type of microplastic exposure measurement (local, transported, or transported from grids over 200km away). "Log seafood exporters' microplastics" is the weighted average of the coastal microplastics concentrations in the seafood exporting countries that the birth state imports from. The weights are determined by the country-year-month seafood importing value. All regressions control for GADM admin level-1 subcountry region fixed effects and country by month-of-sample fixed effects. Standard errors are clustered at the admin level-1 level. *: $p < 0.10$; **: $p < 0.05$; ***: $p < 0.01$.

# Technical Appendix: Ocean Current Modeling Details

To build a matrix that summarizes monthly long distance current flow intensities from ocean grids to coastal locations of interest, we built a model that evaluates the intensity of the current between a sender grid and a receiver city at a given date. Beginning from a particular day and grid, the model constructs streamlines by sequentially following the current's speed and direction on a daily basis.

To define the grids used as senders, we estimated which distance in terms of polar coordinates $res$ was corresponding to a distance of 250 kilometers along the equator. We then used as sender locations the intersections of the grid with a cell size $res$ which were within the latitudes where the microplastics concentrations data were available.

The input current data (current direction and speed information, i.e., vectors) are at a grid-day level, the grid having a precision of 0.25 in terms of polar coordinates.

More precisely, we can illustrate how the algorithm works for a sender grid $i$ when starting at a given day $d$. We initialize the step $t$ by $t = 0$ and the position $p_t$ by the position of the sender grid:

- At step $t$ , we extract from the input data the current direction and speed information at day $d + t$, which is a vector grid with precision 0.25 as indicated previously. Given that $p_t$ is not exactly at an intersection of the grid, linear interpolation is used to approximate the current vector at position $p_t$. Let's note that current vector $c_t = (u_t, v_t)$.

- We look for potential receiver cities within a disk of radius $rad_t$ from $p_t$. For each city $r$ found, we define raw downstream intensity score as:

$$Current_{i \to r, d, t} = \exp\{-\alpha \cdot rad_t - \beta \cdot |\theta|_{i \to r, d, t} - \gamma \cdot dist_{i \to r, d, t}\}$$

where $\alpha, \beta, \gamma$ are positive parameters.

The <u>first</u> component is the search radius at step $t$  ($rad_t$), which captures general decreases of downstream intensity oversteps. We increase the search radius by $0.05$ at each step, which represents about 6km at the equator latitudes, to capture both the uncertainty in the streamline computation and the dispersion of microplastics in the ocean. The initial search radius $rad_0$ has a value of $1$ which represents about 111km at the equator.

The <u>second</u> component enables to assign higher intensity to receiver cities that sit closer to the exactly-downstream direction of the sender grid. More precisely, let's note $l_{i \to r}$ the vector between sender grid $i$ and receiver city $r$. The sense of the vector is not important. Let's note $v_t$ the vector defined as $v_t =$ $(v_t, -u_t)/||c_t||$. Thus, it is a normal vector to the current vector $c_t$ that has a norm of 1. Then, we define the absolute scalar product $|\theta|_{i \to r,d,t} = v_t . l_{i \to r}$. The higher this term, the closer $l_{i \to r}$ is to be perpendicular to $c_t$. Since the higher $|\theta|_{i \to r,d,t}$, the lower the score $Current_{i \to r,d,t}$, and thus the aim of this term is to penalize cities that are less impacted by the current streamlines because they are not in the exactly-downstream direction from the current streamline at that step. It is also important to note that $v_t$ is normalized to avoid seeing lower scores when the speed of the current is higher. However, $l_{i \to r}$ is not normalized to penalize cities that are further from the sender grid.

The <u>third</u> component is simply the distance between the sender and the current position $p_t$ ($dist_{i \to r,d,t}$), which captures geographic decay. This component is inspired by <u>Phillips et al. (2021)</u> that also uses an exponential decay with the distance from the sender to model dispersion of air pollution. This term also aims to penalize cities further from $p_t$ but gives more flexibility to the formula by decorrelating the decay in terms of angle (second term) and the one in terms of distance.

We assume $Current_{i \to r,d,t}$ to be zero if $d_{i \to r,d,t} > rad_t$ (i.e., if receiver city lies outside of the search radius at step $t$) or if $\theta_{i \to r,d,t} > 0.4$ radian (i.e., if the receiver city is not obviously in the downstream direction from the current streamline at that step). We choose parameter values $\{\alpha, \beta, \gamma\} = \{0.8, 0.49, 0.23\}$. These coefficients are chosen empirically so that the function that attributes current scores over 90 days is approximately continuous. For that purpose, we used visualisations consisting in heatmaps that simulate the current scores values not only for cities of interest but for all points of the map for different days and different sender grids. Examples of those heatmaps showing the approximate continuity of the current score function for the final value of the parameters can be seen on the figure below.

- If $t < 89$, coefficients need to be updated for step $t + 1$. We increase $rad_t$ as described previously by $0.05$ to obtain $rad_{t+1}$. We update $p_t = (x_t, y_t)$ by following the local direction and speed of the current i.e. using $c_t$ :

$$x_{t+1} = x_t + 24 * 3600 * u_t / dist_m((x_t, y_t), (x_t + 1, y_t))$$

$$y_{t+1} = y_t + 24 * 3600 * v_t / dist_m((x_t, y_t), (x_t + 1, y_t))$$

$$p_{t+1} = (x_{t+1}, y_{t+1})$$

To understand those expressions, we must consider that $x$ and $y$ coordinates are in degrees while the vectors' coordinates $u$ and $v$ are in m/s. The distance (positive or negative) in meters crossed by the current in 24 hours is of $d_{m,x} = 24 * 3600 * u_t$ along the x-axis and $d_{m,y} = 24 * 3600 * v_t$ along the y-axis. To obtain an approximation of the distance $d_p$ crossed in polar coordinates corresponding to a distance $d_m$ in meters, we use a cross product : if a delta of 1 degree in longitude at the latitude $y_t$ represents $dist_m((x_t, y_t), (x_t + 1, y_t))$ meters, then, an approximation of $d_p$ is $d_p \approx 1 * d_m / dist_m((x_t, y_t), (x_t + 1, y_t))$. Thus,

$$d_{p,x} \approx 1 * d_{m,x} / dist_m((x_t, y_t), (x_t + 1, y_t))$$

$$d_{p,y} \approx 1 * d_{m,y} / dist_m((x_t, y_t), (x_t + 1, y_t))$$

hence the expression of $x_{t+1} = x_t + d_{p,x}$ and $y_{t+1} = y_t + d_{p,y}$.

If point $p_{t+1}$ lies outside of the convex hull of points that represent the position of local current vectors, we stop the algorithm. Indeed, that convex hull represents the ocean area around $p_t$. Thus, if $p_{t+1}$ lies outside of that polygon, it means that it is inland and the streamline must be stopped at $p_t$.

Otherwise, we can proceed to step $t + 1$.

Starting from each particular sender grid and day of the period April 2017 to September 2018, we iterate the procedure for 90 steps (i.e., approximately three months) so for $t = 0$ to $t = 89$.

**Examples of current indexes heat maps for two senders with a circular current on the left and a unidirectional current on the right**

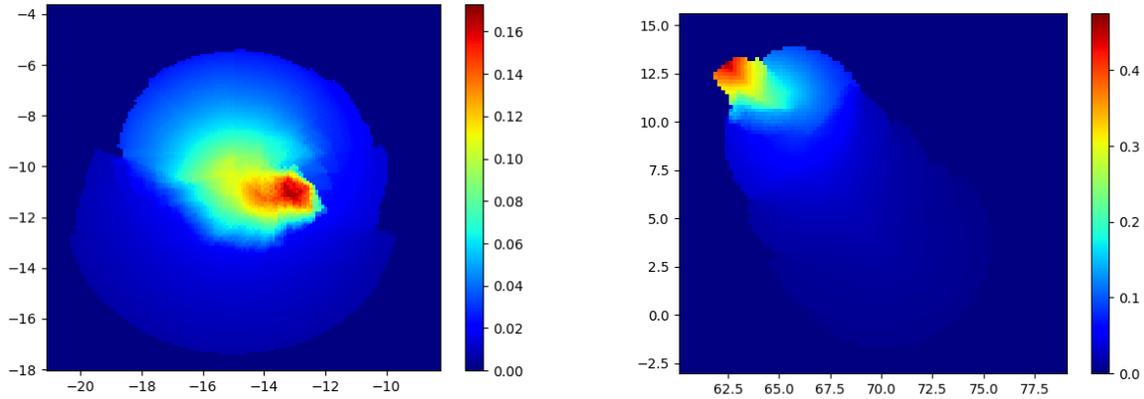

After the computation, we have a set of current scores $Current_{i \rightarrow r,d,t}$ that need to be aggregated at a day level, which means that we want to have a single coefficient for a given tuple (sender grid $i$, receiver city $r$, date of arrival $d'$). For a given index $Current_{i \rightarrow r,d,t}$, the date of arrival $d'$ is the sum of the delay in days from the emission at the sender grid i.e. step $t$ and of the date of emission $d : d' = t + d$. The downstream intensity score aggregated at a day-level for tuple (sender grid $i$, receiver city $r$, date of arrival $d'$) is:

$$Current_{i \rightarrow r,d'} = \sum_{d+t=d'} Current_{i \rightarrow r,d,t}$$

After that, the second aggregation step is at a month-level to make the size of the regression dataset manageable in the econometric analysis. This aggregation consists in computing average intensity scores for each tuple (sender grid $i$, receiver city $r$, month of arrival $m$):

$$\text{Current}_{i \to r, m} = \text{Average}_{d' \in m} \text{Current}_{i \to r, d'}$$

At the end, we record every pair (sender grid $i$, receiver city $r$) among the monthly aggregated intensity scores. For each pair (sender grid $i$, receiver city $r$), when there is no intensity score found for a given month $m$ of the period of interest (1998-2021), we add the monthly aggregated intensity score $\text{Current}_{i \to r, m} = 0$. The final matrix containing monthly intensity scores should therefore present $n$ rows per couple (sender grid $i$, receiver city $r$) where $n = 15$ is the number of months in the period (July 2017-September 2018). Indeed, since there can be up to 90 days between the emission of a streamline at a ocean grid and its reception at a coastal location, only months that begin at least 90 days after the beginning of the period (April 2017) are complete. Note that not every pair (grid, receiver city) would be in the matrix. Indeed, if current "originating" from a grid $i$ has not reached city $r$ within 90 days, for each day of the period as starting date, then no scores will be associated to couple (sender grid $i$, receiver city $r$) in the summary matrix.